\documentclass[reqno,11pt]{amsart}
\usepackage{graphicx}
\usepackage{slashed}
\usepackage{amscd}
\usepackage{tensor}
\usepackage{amsmath,amssymb,mathabx,bm}
\usepackage{esint}
\usepackage{dsfont}
\usepackage{enumitem}
\usepackage{accents}
\usepackage{epsfig}
\usepackage{pst-grad} 
\usepackage{pst-plot} 
\usepackage[mathscr]{eucal}
\definecolor{labelkey}{gray}{.65}
\newcommand{\bitem}{\begin{itemize}[leftmargin=2em]}
\newcommand{\eitem}{\end{itemize}}

\numberwithin{equation}{section}
\allowdisplaybreaks[1]
\title[ Comparing Modified Measures and Causal Fermion Systems]{Modified Measures as an Effective Theory for Causal Fermion Systems}
\author[F.\ Finster]{Felix Finster}
\address{Fakult\"at f\"ur Mathematik \\ Universit\"at Regensburg \\ D-93040 Regensburg \\ Germany}
\email{finster@ur.de}

\author[E. Guendelman]{Eduardo Guendelman}
\address{Department of Physics, Ben Gurion University\\ Be'er Sheva \\ Israel}
\email{guendel@bgu.ac.il}

\author[C.F.\ Paganini]{Claudio F. Paganini$^*$ \\ \\ March / September 2023}
\address{$^*$Fakult\"at f\"ur Mathematik \\ Universit\"at Regensburg \\ D-93040 Regensburg \\ Germany}

\email{claudio.paganini@ur.com}

\newtheorem{Def}{Definition}[section]

\newcommand{\Thanks}{\vspace*{.5em} \noindent \thanks}
\newcommand{\beq}{\begin{equation}}
\newcommand{\eeq}{\end{equation}}

\renewcommand{\L}{\mathcal{L}}
\renewcommand{\H}{\mathscr{H}}
\newcommand{\la}{\langle}
\newcommand{\ra}{\rangle}
\newcommand{\Lin}{\text{\rm{L}}}
\newcommand{\N}{\mathbb{N}}
\DeclareMathOperator{\supp}{supp}
\newcommand{\C}{\mathbb{C}}
\newcommand{\Sact}{{\mathcal{S}}}

\newcommand{\R}{{\mathord{\mathbb R}}}
\newcommand{\tr}{\mbox{tr}}
\newcommand{\F}{\mathscr{F}}

\newcommand{\Pdd}{\slashed{\partial}}
\newcommand*{\myfont}{\fontfamily{ppl}\selectfont}		
\DeclareFontFamily{OT1}{rsfso}{}
\DeclareFontShape{OT1}{rsfso}{m}{n}{ <-7> rsfso5 <7-10> rsfso7 <10-> rsfso10}{}
\DeclareMathOperator{\Sl}{\prec\!}
\DeclareMathOperator{\Sr}{\!\succ}
\DeclareFontFamily{OT1}{rsfso}{}
\DeclareFontShape{OT1}{rsfso}{m}{n}{ <-7> rsfso5 <7-10> rsfso7 <10-> rsfso10}{}
\DeclareMathAlphabet{\mycal}{OT1}{rsfso}{m}{n}
\newcommand{\scrM}{\mycal M}
\newcommand{\s}{\mathfrak{s}}

\newcommand{\x}{{\textit{\myfont x}}}

\DeclareFontFamily{OT1}{rsfso}{}
\DeclareFontShape{OT1}{rsfso}{m}{n}{ <-7> rsfso5 <7-10> rsfso7 <10-> rsfso10}{}
\DeclareMathAlphabet{\mycal}{OT1}{rsfso}{m}{n}

\begin{document}

\maketitle

\begin{abstract} 
We compare the structures of the theory of causal fermion systems (CFS), an approach to unify quantum theory with general relativity (GR), with those of modified measure theories (MMT), which are a set of modified gravity theories. 
Classical spacetimes with MMT can be obtained as the continuum limit of a CFS.
This suggests that MMT could serve as effective descriptions of modifications to GR implied by CFS. The goal is to lay the foundation for future research on exploring which MMTs are consistent with the causal action principle of CFS. 

\end{abstract}

\tableofcontents

\section{Introduction } 
This paper is part of a series of papers comparing the structures and ideas of different approaches to fundamental physics that was started with~\cite{ethcfs} and will be continued in~\cite{cfstdqg}. Each of these papers contains an introduction to the theories under consideration including an overview of their accomplishments. These overviews serve as a starting point for a reader familiar with one of the approaches to engage with the other approach. The papers proceed with a detailed comparison between these two theories. Ultimately, the goal is to motivate the community to establish an extensive collection of such articles as a sort of ``Rosetta's stone'' for approaches to fundamental physics. The hope is that such a set of dictionaries of ideas helps the exchange across approaches and thereby catalyzes progress in the foundations of physics. This addresses a distinctly different goal from overview articles such as~\cite{loll2022quantum, Mielczarek_2018, deBoer:2824341} or~\cite{addazi2022quantum} (which originated the present work) that try to cover the development across many approaches simultaneously. By focusing on two sets of ideas at a time, a greater level of depth can be covered and more specific issues addressed.

Two of the main challenges in fundamental theoretical  physics are finding a theory unifying general relativity (GR) and the standard model of particle physics and explaining the ``dark'' components of the universe, namely dark matter and dark energy. Assuming dark energy to be given by the cosmological constant, its value in the present universe can be determined from the observed accelerated expansion of the universe~\cite{acceleratinguniverse, acceleratinguniverse2, acceleratinguniverse3, acceleratinguniverse4} and is found to be positive. Explaining the value of the cosmological constant from first principle is an open problem, the so-called \textit{cosmological constant problem}. A naive derivation from quantum field theory leads to a number that differs from the observed value by about $120$ orders of magnitude. This number is often cited as "the worst theoretical prediction in the history of physics"~\cite{DISCREPANCY}. However, more elaborate treatments can lead to values much closer to observation, for an overview of these developments see, e.g. \cite{MARTIN2012566}.

Addressing these challenges, in the present paper we will compare modified measure theories (MMT), an approach to modified theories of gravity, and causal fermion systems (CFS), a novel approach to fundamental physics. 
 
 MMT are a set of modified theories of gravity developed by Guendelman and collaborators~\cite{NGVE,scaleinvtmt, Fieltheory, manymeasures, Benisty:2019jqz, measuresModifiedGravityandCosmology}. 
 Originally, these theories were developed  with the goal to modify the variational principle for gravity in such a way that the equations of motion become invariant under the addition of a constant to the  Lagrangian.  This puts gravity conceptually on the same footing as the fields of the standard model which satisfy this symmetry naturally. If one considers the cosmological constant to be an effective description for the energy of the vacuum state in quantum field theory, then this symmetry corresponds to the old wisdom from classical mechanics that one can never measure absolute energy levels but only relative ones.
 
 Thereby, MMT are one possible solution to the cosmological constant problem. In the simplest versions of MMT, the cosmological constant~$\Lambda$ becomes an integration constant to be determined by observation. 
 In the second order formalism, the measure density introduces an additional degree of freedom consisting of a scalar field. Variations of this scheme with two or more measures have been used successfully to model inflation, for a review see~\cite{measuresModifiedGravityandCosmology}\footnote{The attentive reader might notice that additional measures provide more freedom in model building. This freedom manifests in the number of constants of integration. These additional parameters of course facilitate the model-to-data fit. The value of these models lies in the fact that they demonstrate what is possible in terms of phenomenology, could one derive such a model from more fundamental considerations.}.   In this case, the integration constants do not only allow for a shift in the cosmological constant, but also determine the shape of the scalar field potentials. In particular, they determine the level of the flat sectors that could serve to describe the inflationary period as well as the sector responsible for the late slow inflation phase of the universe.

The theory of CFS was developed by Finster and collaborators~\cite{cfs, website} as a new approach to unify GR with the standard model of particle physics. It has been worked out in details in perturbations around Minkowski space and has proven viable in general classical spacetimes. In particular, CFS succeeds in deriving the three generations of fermions in the standard model and in resolving the hierarchy problem. The latter is resolved by the fact that the standard model interactions/gauge fields are obtained at leading order in the expansion with respect to the regularization length~$\varepsilon$ while the coupling of matter to gravity only comes in at the next-to-next-to-leading-order in an expansion in powers of this regularization length\footnote{In causal fermion systems one has to introduce a regularization length~$\varepsilon$ for the various structures to be mathematically well-defined. One then studies the limit where~$\varepsilon\searrow0$. In the context of physics the regularization length can be thought of as the Planck length~$l_p$. }.

In recent years also the quantum field theory limit for the standard model fields has been worked out~\cite{qft, fockentangle, finster2018complex, fockfermionic}. This completes the derivation of our well-established theories in suitable limits (except for the Higgs field, for which the dynamics have not yet been worked out).

The focus of current research is on further solidifying the mathematical foundations of the theory and in obtaining novel predictions for open problems. In this direction it was recently shown~\cite{baryogenesis} that the CFS theory allows for a novel mechanism for baryogenesis. Phenomenological consequences of this mechanism are subject of ongoing research, see, for example ~\cite{cosmo}. The present comparison paper was motivated by the fact, that the mechanism of baryogenesis in~\cite{baryogenesis} in fact requires the introduction of modified measures. 

For more extensive introduction to CFS tailored for physicists see~\cite{dice2014,dice2018,review}, 
whereas~\cite{cfsrev,nrstg} might be a good starting point for mathematicians. For a more gravitational perspective see~\cite{grossmann}.
We also refer the interested reader to the textbooks~\cite{cfs, intro} and the website~\cite{website}. The website is built in a modular way for the interested reader to pick those aspects of the theory they are most interested. 

Some connections between CFS and MMT have already been touched in~\cite{baryogenesis}. 
The present comparison should lay the foundation to exploring these connections in more detail and in particular to launch a systematic investigation as to which of the phenomenological results in MMT can be recovered in the setting of CFS. On the one hand, this would be a major step in connecting CFS to observation/experiment. On the other hand, this would allow for a constraint on MMTs as an effective theory of an underlying fundamental theory. 
 
In an upcoming paper~\cite{cfstdqg} some of the authors will compare the CFS theory with thermodynamic approaches to gravity~\cite{bekenstein1973black,bardeen1973four,hawking1975particle,gibbons1977action,jacobson1995thermodynamics,wald1999gravitation,chirco2010nonequilibrium,jacobson2016entanglement,bueno2017entanglement,parikh2018einstein,svesko2019entanglement,alonso2022thermodynamics,banihashemi2022thermodynamic,jacobson2023partition, requardt2023thermal}  with a particular focus on the work by Padmanabhan~\cite{padmanabhan2011entropy,padmanabhan2014general,padmanabhan2015one,padmanabhan2017cosmic,padmanabhan2017atoms}. 
The underlying motivation behind Padmanabhan's approach is the same as for MMT to resolve the cosmological constant problem. In the above-mentioned paper, we will argue that CFS in a certain sense incorporates both sets of ideas to resolve the cosmological constant problem. The reason to mention the thermodynamic approach here is that it has recently been claimed that the theory of gravity emerging from these thermodynamic considerations is unimodular gravity~\cite{alonso2022thermodynamics}. It is worth noting that a covariant formulation of unimodular gravity is a subset of two-measure-theories (TMT). This connection will be explored in~\cite{gravvariance} where it is shown that unimodular gravity is the subset of canonical ensembles while MMT correspond to grand canonical ensembles. This result is a direct spin-off of the current work demonstrating the value of such efforts to compare approaches. Finally, a similar analysis to the one presented here could be done in the future comparing the present two approaches to the torsional formulation of GR, the so-called teleparallel equivalent of GR (TEGR)~\cite{krvsvsak2016covariant,cai2016f}.  
 
\subsection{Organization}
The paper is organized as follows. In Section~\ref{sec:background} we give a brief introduction to the theory of CFS and MMTs. The core of the paper is Section~\ref{sec:comparison}, where we compare various aspects of the two theories, in particular their fundamental structures (Section~\ref{sec:foundations}) and the equations of motion (Section~\ref{sec:eom}). In Section~\ref{sec:metric} we discuss which new structures are introduced in order
to replace the metric and how the different tasks which the metric fulfills in the standard second order formulation of GR are reassigned to those structures at the fundamental level. Finally, in Section~\ref{sec:conclusion} we give a brief conclusion and outlook on future research.

\section{The Theories}\label{sec:background} 
In this section we will give a basic introduction to CFS and MMT. The main purpose of this section is to serve as a starting point for a reader familiar with one of the approaches to engage with the other approach. To that end the overviews introduce the basic ideas of both theories, as well as their key results. The focus is on those mathematical and physical concepts for which something interesting can be said in the next section. Therefore we will keep these introductions relatively short and provide a guide to more in depth literature, both introductory and on current research, for readers interested in engaging with one of the approaches at a deeper level.

 \subsection{Causal Fermion Systems} \label{sec:cfs}
 The central postulate of the CFS theory is the \textit{causal action principle}, from which the field equations of GR and the SM can be derived in the continuum limit (see~\cite{cfs}). The theory thus provides a unification of the weak, the strong and the electromagnetic forces with gravity on the level of second-quantized Dirac fields
 coupled to classical gauge fields.
 Moreover, the approach has led to concise notions of ``quantum spacetime'' and ``quantum geometry''
(see~\cite{rrev, lqg}) and the quantum field theory limit has recently been established rigorously (see~\cite{fockbosonic, fockfermionic,fockentangle}). 

As CFS introduces an entirely new language to encode our physical words, we begin with a few preliminary considerations to develop an understanding, for how the basic concepts of the theory are motivated. When introducing the theory, we explain how its concepts relate to more familiar structures in physics.

\subsubsection{Preliminary considerations}

\begin{figure}[t]
\psset{xunit=.5pt,yunit=.5pt,runit=.5pt}
\begin{pspicture}(525.68616972,189.66897799)
{
\newrgbcolor{curcolor}{0.79607844 0.79607844 0.79607844}
\pscustom[linestyle=none,fillstyle=solid,fillcolor=curcolor]
{
\newpath
\moveto(52.52530394,173.45811185)
\lineto(494.46248693,172.64651122)
\lineto(337.82064756,7.14107878)
\lineto(210.50249197,7.14107878)
\closepath
}
}
{
\newrgbcolor{curcolor}{0.80000001 0.80000001 0.80000001}
\pscustom[linewidth=0.99999871,linecolor=curcolor]
{
\newpath
\moveto(52.52530394,173.45811185)
\lineto(494.46248693,172.64651122)
\lineto(337.82064756,7.14107878)
\lineto(210.50249197,7.14107878)
\closepath
}
}
{
\newrgbcolor{curcolor}{0 0 0}
\pscustom[linewidth=2.00000125,linecolor=curcolor]
{
\newpath
\moveto(506.78631685,8.41585295)
\lineto(8.4,6.52802429)
\lineto(8.80885039,161.36545295)
}
}
{
\newrgbcolor{curcolor}{0 0 0}
\pscustom[linestyle=none,fillstyle=solid,fillcolor=curcolor]
{
\newpath
\moveto(503.96512319,14.00521027)
\lineto(519.38623431,8.46357996)
\lineto(504.00754719,2.80528364)
\curveto(506.79517286,6.07786827)(506.77751387,10.73983773)(503.96512319,14.00521027)
\closepath
}
}
{
\newrgbcolor{curcolor}{0 0 0}
\pscustom[linestyle=none,fillstyle=solid,fillcolor=curcolor]
{
\newpath
\moveto(3.20147301,158.58024781)
\lineto(8.84212077,173.96541688)
\lineto(14.40144094,158.55067413)
\curveto(11.1468437,161.35927945)(6.48485705,161.37158949)(3.20147301,158.58024781)
\closepath
}
}
{
\newrgbcolor{curcolor}{0 0 0}
\pscustom[linewidth=1.49999998,linecolor=curcolor]
{
\newpath
\moveto(209.9585348,7.6574076)
\lineto(53.06926866,172.94177547)
}
}
{
\newrgbcolor{curcolor}{0 0 0}
\pscustom[linewidth=1.49999998,linecolor=curcolor]
{
\newpath
\moveto(338.36463496,7.6574076)
\lineto(495.25392,172.94177547)
}
}
{
\newrgbcolor{curcolor}{0 0 0}
\pscustom[linewidth=3.99999979,linecolor=curcolor]
{
\newpath
\moveto(210.50249575,7.14107122)
\lineto(337.82064756,7.14107122)
}
\rput[bl](260,20){$\psi_0$}
\rput[bl](20,160){$t$}
\rput[bl](505,20){$\vec{x}$}
}
\end{pspicture}
\caption{Causal propagation of a wave function.}
\label{figmmt1}
\end{figure}
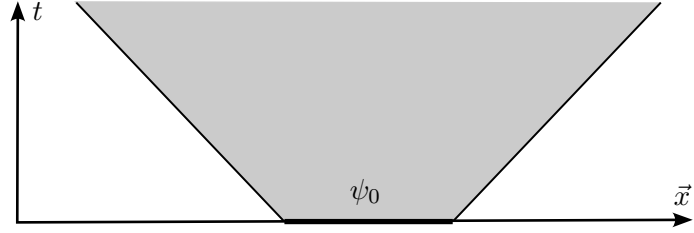

To build up to the structures used to represent the physical world in CFS we start with some considerations that should be familiar to most readers. Suppose we have a quantum mechanical wave function $\Psi$ describing a particle satisfying the Klein-Gordon equation
\begin{equation}
    (\nabla^\mu\nabla_\mu -m^2)\psi (\x) =0 \:.
\end{equation}
Now suppose further that we have only access to the information contained in the probability density $|\psi(\x)|^2$ of the wave function. The question then is, given this information, what can we say about the classical spacetime? First let the wave function $\Psi$ be a solution evolved from compactly supported initial data as illustrated in Fig.~\ref{figmmt1}.
 Then the finite speed of propagation guarantees that the probability density $|\psi(\x)|^2$ vanishes outside the causal future of the support of the initial data. Therefore, the support of the probability density $|\psi(\x)|^2$ contains some information about the causal structure of the classical spacetime. 
 
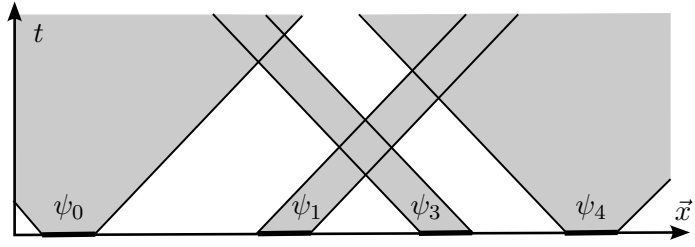
\begin{figure}[t]
\psset{xunit=.5pt,yunit=.5pt,runit=.5pt}
\begin{pspicture}(516.58620279,180.56370532)
{
\newrgbcolor{curcolor}{0.79607844 0.79607844 0.79607844}
\pscustom[linestyle=none,fillstyle=solid,fillcolor=curcolor]
{
\newpath
\moveto(268.69745386,170.15284548)
\lineto(501.43032945,171.2699491)
\lineto(501.34642394,46.53398186)
\lineto(463.48636724,7.66710438)
\lineto(422.41223811,6.25643729)
\closepath
}
}
{
\newrgbcolor{curcolor}{0.80000001 0.80000001 0.80000001}
\pscustom[linewidth=0.99999871,linecolor=curcolor]
{
\newpath
\moveto(268.69745386,170.15284548)
\lineto(501.43032945,171.2699491)
\lineto(501.34642394,46.53398186)
\lineto(463.48636724,7.66710438)
\lineto(422.41223811,6.25643729)
\closepath
}
}
{
\newrgbcolor{curcolor}{0.79607844 0.79607844 0.79607844}
\pscustom[linestyle=none,fillstyle=solid,fillcolor=curcolor]
{
\newpath
\moveto(194.94964535,169.57664139)
\lineto(157.08248315,170.19181619)
\lineto(313.7243263,4.68638375)
\lineto(351.9130885,4.25478438)
\closepath
}
}
{
\newrgbcolor{curcolor}{0.80000001 0.80000001 0.80000001}
\pscustom[linewidth=0.99999871,linecolor=curcolor]
{
\newpath
\moveto(194.94964535,169.57664139)
\lineto(157.08248315,170.19181619)
\lineto(313.7243263,4.68638375)
\lineto(351.9130885,4.25478438)
\closepath
}
}
{
\newrgbcolor{curcolor}{0.79607844 0.79607844 0.79607844}
\pscustom[linestyle=none,fillstyle=solid,fillcolor=curcolor]
{
\newpath
\moveto(6.49712126,170.15284548)
\lineto(223.74078992,170.15284548)
\lineto(66.22308661,5.85330532)
\lineto(27.80870173,2.77807761)
\lineto(6.81657071,28.60848013)
\closepath
}
}
{
\newrgbcolor{curcolor}{0.80000001 0.80000001 0.80000001}
\pscustom[linewidth=0.99999871,linecolor=curcolor]
{
\newpath
\moveto(6.49712126,170.15284548)
\lineto(223.74078992,170.15284548)
\lineto(66.22308661,5.85330532)
\lineto(27.80870173,2.77807761)
\lineto(6.81657071,28.60848013)
\closepath
}
}
{
\newrgbcolor{curcolor}{0.79607844 0.79607844 0.79607844}
\pscustom[linestyle=none,fillstyle=solid,fillcolor=curcolor]
{
\newpath
\moveto(346.49940661,170.67854375)
\lineto(385.93114205,170.76732485)
\lineto(229.28930268,5.26189241)
\lineto(188.52221858,4.3615069)
\closepath
}
}
{
\newrgbcolor{curcolor}{0.80000001 0.80000001 0.80000001}
\pscustom[linewidth=0.99999871,linecolor=curcolor]
{
\newpath
\moveto(346.49940661,170.67854375)
\lineto(385.93114205,170.76732485)
\lineto(229.28930268,5.26189241)
\lineto(188.52221858,4.3615069)
\closepath
}
}
{
\newrgbcolor{curcolor}{0 0 0}
\pscustom[linewidth=2.00000125,linecolor=curcolor]
{
\newpath
\moveto(503.98631055,5.61058028)
\lineto(5.5999937,3.72275162)
\lineto(5.5999937,167.96370532)
}
}
{
\newrgbcolor{curcolor}{0 0 0}
\pscustom[linestyle=none,fillstyle=solid,fillcolor=curcolor]
{
\newpath
\moveto(501.16511689,11.1999376)
\lineto(516.58622801,5.65830729)
\lineto(501.20754089,0.00001097)
\curveto(503.99516656,3.2725956)(503.97750757,7.93456506)(501.16511689,11.1999376)
\closepath
}
}
{
\newrgbcolor{curcolor}{0 0 0}
\pscustom[linestyle=none,fillstyle=solid,fillcolor=curcolor]
{
\newpath
\moveto(-0.00000979,165.16370358)
\lineto(5.5999937,180.56371318)
\lineto(11.19999719,165.16370358)
\curveto(7.93799516,167.96370532)(3.27599225,167.96370532)(-0.00000979,165.16370358)
\closepath
}
}
{
\newrgbcolor{curcolor}{0 0 0}
\pscustom[linewidth=1.49999998,linecolor=curcolor]
{
\newpath
\moveto(312.49049953,5.7360606)
\lineto(155.60122961,171.02042847)
}
}
{
\newrgbcolor{curcolor}{0 0 0}
\pscustom[linewidth=1.49999998,linecolor=curcolor]
{
\newpath
\moveto(66.30753638,4.35214123)
\lineto(223.1968252,169.6365091)
}
}
{
\newrgbcolor{curcolor}{0 0 0}
\pscustom[linewidth=3.99998762,linecolor=curcolor]
{
\newpath
\moveto(26.42963906,3.83580863)
\lineto(66.62276787,4.33579855)
}
}
{
\newrgbcolor{curcolor}{0 0 0}
\pscustom[linewidth=3.99998762,linecolor=curcolor]
{
\newpath
\moveto(189.73374236,4.7029154)
\lineto(229.92687118,5.20290532)
}
}
{
\newrgbcolor{curcolor}{0 0 0}
\pscustom[linewidth=3.99998762,linecolor=curcolor]
{
\newpath
\moveto(421.68123969,5.16727572)
\lineto(461.87439874,5.66726564)
}
}
{
\newrgbcolor{curcolor}{0 0 0}
\pscustom[linewidth=3.99998762,linecolor=curcolor]
{
\newpath
\moveto(311.55567118,4.96973115)
\lineto(351.74879244,5.46972107)
}
}
{
\newrgbcolor{curcolor}{0 0 0}
\pscustom[linewidth=1.49999998,linecolor=curcolor]
{
\newpath
\moveto(27.23871874,3.2655384)
\lineto(4.70387528,29.61521777)
}
}
{
\newrgbcolor{curcolor}{0 0 0}
\pscustom[linewidth=1.5090746,linecolor=curcolor]
{
\newpath
\moveto(188.8727622,4.91506784)
\lineto(347.13878929,170.75044548)
}
}
{
\newrgbcolor{curcolor}{0 0 0}
\pscustom[linewidth=1.49999998,linecolor=curcolor]
{
\newpath
\moveto(229.66205102,5.46924485)
\lineto(386.55132472,170.75361273)
}
}
{
\newrgbcolor{curcolor}{0 0 0}
\pscustom[linewidth=1.49999998,linecolor=curcolor]
{
\newpath
\moveto(461.29514835,5.93360517)
\lineto(501.82676409,46.00060359)
}
}
{
\newrgbcolor{curcolor}{0 0 0}
\pscustom[linewidth=1.49999998,linecolor=curcolor]
{
\newpath
\moveto(423.00097512,5.46924485)
\lineto(266.11172787,170.75361273)
}
}
{
\newrgbcolor{curcolor}{0 0 0}
\pscustom[linewidth=1.49999998,linecolor=curcolor]
{
\newpath
\moveto(352.49812157,4.91034721)
\lineto(195.60889323,170.19471509)
}
\rput[bl](35,15){$\psi_0$}
\rput[bl](215,15){$\psi_1$}
\rput[bl](305,15){$\psi_3$}
\rput[bl](430,15){$\psi_4$}
\rput[bl](20,150){$t$}
\rput[bl](505,15){$\vec{x}$}
}
\end{pspicture}
\caption{Probing with many wave functions.}
\label{figmmt4}
\end{figure}

Of course, there is only a limited amount of information which one can extract from a single wave function. However, if instead we probe classical spacetime with many wave functions, as illustrated in Fig.~\ref{figmmt4}, the situation is different.
If we aggregate over all wave functions evolved from compact initial data, then we can extract the complete causal structure of the classical spacetime. This fixes the metric up to a conformal factor~\cite{hawking1976new,malament1977class}, i.e.,\ up to the volume form / volume density.

\begin{figure}[t]
\psset{xunit=.5pt,yunit=.5pt,runit=.5pt}
\begin{pspicture}(525.68616972,189.66897799)
{
\newrgbcolor{curcolor}{0.79607844 0.79607844 0.79607844}
\pscustom[linestyle=none,fillstyle=solid,fillcolor=curcolor]
{
\newpath
\moveto(186.78460346,174.39493201)
\lineto(146.95739339,173.52453705)
\lineto(304.91785701,7.20749642)
\lineto(343.14894992,7.72464539)
\closepath
}
}
{
\newrgbcolor{curcolor}{0.80000001 0.80000001 0.80000001}
\pscustom[linewidth=0.99999871,linecolor=curcolor]
{
\newpath
\moveto(186.78460346,174.39493201)
\lineto(146.95739339,173.52453705)
\lineto(304.91785701,7.20749642)
\lineto(343.14894992,7.72464539)
\closepath
}
}
{
\newrgbcolor{curcolor}{0.79607844 0.79607844 0.79607844}
\pscustom[linestyle=none,fillstyle=solid,fillcolor=curcolor]
{
\newpath
\moveto(219.21427654,173.45811185)
\lineto(274.70850898,173.52671028)
\curveto(208.15830803,131.00071343)(158.93693858,62.48147311)(101.05116094,6.95884886)
\lineto(62.84993008,6.78782524)
\closepath
}
}
{
\newrgbcolor{curcolor}{0.80000001 0.80000001 0.80000001}
\pscustom[linewidth=0.99999871,linecolor=curcolor]
{
\newpath
\moveto(219.21427654,173.45811185)
\lineto(274.70850898,173.52671028)
\curveto(208.15830803,131.00071343)(158.93693858,62.48147311)(101.05116094,6.95884886)
\lineto(62.84993008,6.78782524)
\closepath
}
}
{
\newrgbcolor{curcolor}{0 0 0}
\pscustom[linewidth=2.00000125,linecolor=curcolor]
{
\newpath
\moveto(404.79584126,6.9559991)
\lineto(8.4,6.52802429)
\lineto(8.80885039,161.36545295)
}
}
{
\newrgbcolor{curcolor}{0 0 0}
\pscustom[linestyle=none,fillstyle=solid,fillcolor=curcolor]
{
\newpath
\moveto(401.98979502,12.55297626)
\lineto(417.39584177,6.96960288)
\lineto(402.00188727,1.35297581)
\curveto(404.79836552,4.617999)(404.79333212,9.27999919)(401.98979502,12.55297626)
\closepath
}
}
{
\newrgbcolor{curcolor}{0 0 0}
\pscustom[linestyle=none,fillstyle=solid,fillcolor=curcolor]
{
\newpath
\moveto(3.20147301,158.58024781)
\lineto(8.84212077,173.96541688)
\lineto(14.40144094,158.55067413)
\curveto(11.1468437,161.35927945)(6.48485705,161.37158949)(3.20147301,158.58024781)
\closepath
}
}
{
\newrgbcolor{curcolor}{0 0 0}
\pscustom[linewidth=1.49999998,linecolor=curcolor]
{
\newpath
\moveto(304.59402709,7.20749642)
\lineto(145.9438148,173.00820067)
}
}
{
\newrgbcolor{curcolor}{0 0 0}
\pscustom[linewidth=1.49999998,linecolor=curcolor]
{
\newpath
\moveto(101.59512567,7.47518524)
\lineto(181.36025953,90.12086713)
}
}
{
\newrgbcolor{curcolor}{0 0 0}
\pscustom[linewidth=3.99998762,linecolor=curcolor]
{
\newpath
\moveto(61.84010079,7.14107122)
\lineto(101.83100598,7.6413824)
}
}
{
\newrgbcolor{curcolor}{0 0 0}
\pscustom[linewidth=1.50047261,linecolor=curcolor]
{
\newpath
\moveto(62.56143496,8.15771878)
\lineto(219.4507389,173.44208665)
}
}
{
\newrgbcolor{curcolor}{0 0 0}
\pscustom[linewidth=3.99998762,linecolor=curcolor]
{
\newpath
\moveto(303.37705701,7.20749642)
\lineto(344.31969638,7.20749642)
}
}
{
\newrgbcolor{curcolor}{0 0 0}
\pscustom[linewidth=1.49999998,linecolor=curcolor]
{
\newpath
\moveto(345.50616567,6.70458492)
\lineto(188.61680504,173.52453705)
}
}
{
\newrgbcolor{curcolor}{0.60000002 0.60000002 0.60000002}
\pscustom[linestyle=none,fillstyle=solid,fillcolor=curcolor]
{
\newpath
\moveto(194.966689,73.21791798)
\curveto(194.966689,60.10930335)(170.80770487,49.48267155)(141.00608768,49.48267155)
\curveto(111.20447048,49.48267155)(87.04548635,60.10930335)(87.04548635,73.21791798)
\curveto(87.04548635,86.32653261)(111.20447048,96.95316441)(141.00608768,96.95316441)
\curveto(170.80770487,96.95316441)(194.966689,86.32653261)(194.966689,73.21791798)
\closepath
}
}
{
\newrgbcolor{curcolor}{0.60000002 0.60000002 0.60000002}
\pscustom[linewidth=2.4999988,linecolor=curcolor]
{
\newpath
\moveto(194.966689,73.21791798)
\curveto(194.966689,60.10930335)(170.80770487,49.48267155)(141.00608768,49.48267155)
\curveto(111.20447048,49.48267155)(87.04548635,60.10930335)(87.04548635,73.21791798)
\curveto(87.04548635,86.32653261)(111.20447048,96.95316441)(141.00608768,96.95316441)
\curveto(170.80770487,96.95316441)(194.966689,86.32653261)(194.966689,73.21791798)
\closepath
}
}
{
\newrgbcolor{curcolor}{0 0 0}
\pscustom[linewidth=1.49999998,linecolor=curcolor]
{
\newpath
\moveto(180.86804787,90.6417087)
\curveto(183.85080567,94.1623802)(186.8901052,97.63514618)(189.98438173,101.05821894)
\curveto(215.20479496,128.95850618)(244.09594583,153.53668319)(275.67792,173.95910335)
}
\rput[bl](85,15){$\psi_0$}
\rput[bl](300,15){$\phi_0$}
\rput[bl](20,160){$t$}
\rput[bl](405,20){$\vec{x}$}
\rput[bl](102,67){EM field}
}
\end{pspicture}
\caption{Probing an electromagnetic field.}
\label{figmmt3}
\end{figure}
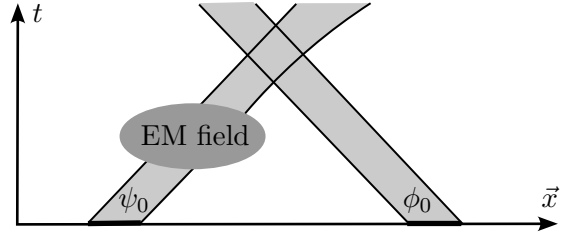

Next, if we allow the wave function to couple to the electromagnetic field, e.g., through minimal coupling
\begin{equation}
    \left((\nabla^\mu+iqA^\mu)(\nabla_\mu +iqA_\mu)-m^2\right)\psi (\x) =0 \:.
\end{equation}
and instead of just the probability density of a single wave function we consider the entire information contained in the correlations $\overline{\psi(\x)}\phi(x)$ of all the wave functions. Then one can expect that this also encodes information about the electromagnetic fields in the classical spacetime, as illustrated in Fig.~\ref{figmmt3}. Therefore there is hope that, increasing the number of wave functions, we can recover both the classical spacetime structure and the matter fields therein. 

Now we go one step further and mathematically formalize the idea of encoding classical spacetime
in a family of wave functions. Consider wave functions, mapping from a classical spacetime into the complex numbers $\psi_1,\dots,\psi_f: \, \mathcal{M} \rightarrow \C$, which are orthonormal vectors in a Hilbert space
\begin{equation}
    \la \psi_k|\psi_l\ra_\H = \delta_{kl}.
\end{equation}
They form an $f$ dimensional Hilbert space. We can now introduce the local correlation operator at a point $\x$, $F(\x): \H \rightarrow \H$
as
\begin{equation}
\left(F(\x)\right)^j_{\hphantom{j}k}=\overline{\psi_j(\x)}\psi_k(\x),
\end{equation}
in terms of matrix elements, or

\begin{equation}
\la \psi|F(\x)\phi\ra_\H=\overline{\psi(\x)}\phi(\x), \qquad \text{ for all } \psi, \phi \in \H,
\end{equation}
in basis invariant form. From the construction it is clear that the operator $F(x)$ is positive semi definite of rank at most one. We have thus constructed a map from the classical spacetime to the set $\mathcal{B}$ of semi definite bounded linear operators\footnote{In this paper we use $\Lin(\H)$ to denote the set of all bounded linear operators on a Hilbert space.} and of rank at most one on a Hilbert space of dimension $f$.
\begin{equation}
    \mathcal{B}:= \{ y \in \Lin(\H) \,|\, y \text{ positive semi-definite of rank at most one}\}
\end{equation}
  This map encodes all the physical information contained in the wave functions. As mentioned before this information does not include the volume measure.
  Therefore, we need to introduce the volume measure as an additional structure.
  We can define a volume measure on the set $\mathcal{B}$ by the push-forward measure defined by
\begin{equation}
  \rho (\Omega):= \mu(F^{-1}(\Omega))= \int_{F^{-1}(\Omega)} \Phi\, \,d^4\x \:.
\end{equation}
We now have all the ingredients at hand needed to define a CFS. However, in that case instead of complex wave functions we will work with sections of a spinor bundle. One consequence of that is that the local correlation operator will no longer be positive semi-definite. They will still be of finite rank but instead with a fixed 
upper bound for the number of positive and negative eigenvalues.

\subsubsection{General Definitions}
We are now ready to introduce the abstract definition of a CFS. It consists of three objects: 
\begin{enumerate}
    \item a Hilbert space~$\H$,
    \item a suitably chosen subset~$\F$ of the bounded linear operators on the Hilbert space, 
    \item and a measure~$\rho$ defined on the Borel $\sigma$-algebra with respect to the norm-topology on the bounded linear operators~$\Lin(\H)$.
\end{enumerate}

\begin{Def} \label{def:cfs}
Let~${(\H, \la .|. \ra_\H)}$ be a Hilbert space. 
Given the parameter {$n \in \N$} (``{spin dimension}''), we let~$\F$ be the set
\begin{align*}
{\F} := \Big\{ \;y &\in \Lin(\H) \text{ with the properties:} \\
&\blacktriangleright\; \text{$y$ is {self-adjoint} and has {finite rank}} \\
&\blacktriangleright\; \text{$y$ has at most~$n$ positive and at most~$n$ negative eigenvalues} \;\Big\} .
\end{align*}
Moreover, let~$\rho$ be a Borel measure on~$\F$.
We refer to~$(\H, \F, \rho)$ as a {\bf{causal fermion system}} (CFS).
\end{Def} 

The operators with exactly~$n$ positive and~$n$ negative eigenvalues constitute the subset~$\F^{\text{reg}}$ of \textit{regular points} in the set~$\F$. One can show~\cite{gaugefix, banach} that the subset~$\F^{\text{reg}}$ is a Banach manifold. Hence, the theory of causal fermion systems is based on the mathematical structure of an \textit{operator manifold}, thereby fusing the underlying structure of GR and quantum theory. The measure then encodes the classical spacetime and all structures therein. 
In order to understand how this comes about, we need to be able to connect these fundamental descriptions to established notions of classical spacetimes and matter in GR. In the following, we will only give a schematic description of how this works. The key tool will be the local correlation map which we introduced for a simplified setting in the preliminary considerations. To begin with, we only introduce it as an abstract map

\begin{align*} 
    F\,[g_{\mu\nu}, A_\mu,\dots]: \mathcal{M}  \qquad&\rightarrow \qquad \;\; \qquad\F \\
    \x \qquad&\rightarrow \qquad F\,[g_{\mu\nu}, A_\mu,\dots](\x).
\end{align*}
The crucial point is that the map allows us to identify points in a classical spacetime~$\mathcal{M}$ with operators in the set~$\F$. The map~$F$ depends on the metric~$g_{\mu\nu}$, as well as the matter fields defined on the classical spacetime under consideration. Hence all these structures are encoded in the map itself. We will explain the construction of this map in Minkowski space in some detail in Section~\ref{sec:continuum} while the precise construction can be found in~\cite{cfs}.
 We can then use the local correlation map to define the measure~$\rho$ of a subset~$\Omega \subset \F$ as the push forward of the measure~$\mu$ of the classical spacetime under the map $F$
\begin{equation}\label{eq:pushforward}
  \rho (\Omega):= \mu(F^{-1}(\Omega))= \int_{F^{-1}(\Omega)} \Phi\, d^4\x.
\end{equation}
Here~$\Phi$ is the density function of the measure on the manifold. If one considers the geometric measure on the manifold then~$\Phi=\sqrt{|g|}$. However the construction of the push forward measure also works for more general measures on the manifold~$\mathcal{M}$. In particular it allows for the measures studied in the context of MMT, which will be introduced in Section~\ref{sec:nrmt}. 

For further considerations, it is convenient to introduce the notation
\begin{align*}
S_{F(\x)} &:= F(\x)(\H) \subset \H &&\hspace*{-2cm} \text{subspace of dimension~$\leq 2n$} \\
\pi_{F(\x)} &: \H \rightarrow S_{F(\x)}\subset \H &&\hspace*{-2cm} \text{orthogonal projection on~$S_{F(\x)}$} \:.
\end{align*}
With that notation at hand, the local correlation map allows us to give an explicit realization of the Hilbert space in the manifold~$\mathcal{M}$ in terms of physical wave functions
\[ \psi^u(\x) = \pi_{F(\x)} u \:. \]
Here~$u$ is a vector in the Hilbert space~$\H$. As a result of this construction, for every point~$\x$,
the physical wave function~$\psi^u(\x)$ is a~$2n$ dimensional vector that is an element in the vector space~$S_{F(\x)}$. One can think of the vector space~$S_{F(\x)}$ as a fiber space attached to a point~$F(\x)$ in the operator manifold. The realization of vectors in the Hilbert space in terms of physical wave functions allows us to formulate a CFS entirely in the language of classical spacetimes and structures therein. In the other direction, the local correlation map allows us to describe any classical configuration of spacetime and matter as a CFS. These classical structures can be thought of as an (approximate) 
effective description. 

With the local correlation map at hand, we can introduce the notion of "spacetime" in the setting of CFS as the support\footnote{The
{\em{support}} of a measure is defined as the complement of the largest open set of measure zero, i.e.,
\[ \supp \rho := \F \setminus \bigcup \big\{ \text{$\Omega \subset \F$ \,\big|\,
$\Omega$ is open and~$\rho(\Omega)=0$} \big\} \:. \]} of the measure, 
\[  M := \supp \rho \:. \]
Hence, in the contrast to GR where the support of the measure is the entire manifold on which the variation principle is defined, in the context of CFS the measure is the fundamental variable and its support is typically only a proper subset of the set~$\F$. In particular, the \textit{dimension of the spacetime in CFS can be much lower} than the dimension of of the set~$\F$ (or the manifold~$\F^{\text{reg}}$ for that matter) which can in principal be infinite. 
Furthermore, despite the subset~$\F^{\text{reg}}$ being a manifold, the spacetime $M$ need not be so, but can in fact also have a \textit{discrete structure}. 
This explains how a single scalar quantity, like the density function of the measure $\rho$, can describe such a plethora of physical systems: By changing the configuration of the metric and matter fields in the local correlation map~$F[g_{\mu\nu}, A_\mu,\dots]$ we map into a different subset of the set~$\F$. This also gives us a different realizations of the Hilbert space in the classical spacetime~$\mathcal{M}$ in terms of physical wave functions\footnote{ A change in the local correlation map identifies a different operator with a particular point in the classical spacetime. This changes the value of the physical wave functions at that point. Therefore, if we vary the measure by changing the local correlation map, we vary the family of all physical wave functions in spacetime. 
In fact, as we will show below, in examples of causal fermion systems describing classical spacetimes we reverse the order of the construction.
Hence, we change the realization of the Hilbert space in terms of physical wave functions to vary the measure by changing the local correlation map. Therefore, in the continuum limit, a variation of the measure can be thought of equivalently as a variation of the family of all physical wave functions. }

To summarize, the points in the set~$M$ are called {\em{spacetime points}}, and the entire set~$M$ is referred to as {\em{spacetime}}.
In order to implement this identification in a compact notation,
in the following we will omit writing the map $F$ explicitly and refer to elements in the set $\F$ by $x$ with $x=F(\x)$ whenever a CFS can be written as the image of a local correlation map from a classical spacetime. Throughout the paper classical spacetime refers to a $(3+1)$-dimensional, time-orientable, Lorentzian manifold while spacetime corresponds to the aforementioned structure in CFS. 

We now proceed to summarize those structures presented in~\cite{cfs}, which will be essential for what follows. In particular we will focus on the fact that starting from the abstract definition of a CFS, one can obtain the usual spacetime structures such as causal relations. To achieve this, given a CFS, one introduces
additional mathematical objects which are {\em{inherent}} in the sense that they
only use information already encoded in the CFS.
Then one shows that these inherent structures correspond to familiar
structures in a classical Lorentzian spacetime, at least in suitable limiting cases.

In order to define a causal structure in spacetime,
we study the spectral properties of the operator product~$xy$.
This operator product is still an operator
of rank at most~$2n$. However, in general it is not symmetric (note that~$(xy)^* = yx \neq xy$ unless the
operators commute) and therefore not an element of~$\F$.
We denote the rank of~$xy$ by~$k \leq 2n$. Counting algebraic multiplicities, we choose~$\lambda^{xy}_1, \ldots, \lambda^{xy}_{k} \in \C$ as all the non-zero eigenvalues and set~$\lambda^{xy}_{k+1}, \ldots, \lambda^{xy}_{2n}=0$.

These eigenvalues give rise to the following notion of a causal structure.
\begin{Def} \label{def:causalstructure}
The points~$x,y \in \F$ are said to be
\[ \left\{ \begin{array}{cll}
\text{{\bf{spacelike}} separated} &\!&  {\mbox{if~$|\lambda^{xy}_j|=|\lambda^{xy}_k|$
for all~$j,k=1,\ldots, 2n$}} \\[0.4em]
\text{{\bf{timelike}} separated} &&{\mbox{if~$\lambda^{xy}_1, \ldots, \lambda^{xy}_{2n}$ are all real}} \\[0.2em]
&& \text{and~$|\lambda^{xy}_j| \neq |\lambda^{xy}_k|$ for some~$j,k$} \\[0.2em]
\text{{\bf{lightlike}} separated} && \text{otherwise}\:.
\end{array} \right. \]
\end{Def} \noindent
More specifically, the points~$x$ and~$y$ are lightlike 
separated if not all the eigenvalues have the same absolute value and if  not all of them are real.
We point out that this notion of causality does not rely on an underlying metric.
But it reduces to the causal structure of Minkowski space or general curved classical spacetimes in certain limiting
cases (for Minkowski space see~\cite[Section~1.2]{cfs} or Section~\ref{sec:continuum};
for curved classical spacetime see~\cite{lqg}). This notion of causality will be explained a bit more later in this section.

Next we want to define a connection. In a classical spacetime the connection is defined pointwise. However, 
as its name implies, it serves the role to connect between the fibres of neighboring points in a classical spacetime, i.e., to encode which elements in the fibers can be identified if we move within the manifold. More precisely, the so-called {\em{spin connection}}~$D_{x,y}$
is a unitary map from~$S_x$ to~$S_y$,
\begin{equation} \label{eq:cfsconnection}
    D_{x,y} \::\: S_y \rightarrow S_x \qquad \text{unitary}\:.
\end{equation}
It can be constructed from the kernel of the fermionic projector~$P(x,y)=\pi_y x$. For details of this construction see~\cite{lqg}. 
The spin connection also gives rise to a corresponding metric connection, being a mapping between corresponding Clifford structures.
The associated curvature is introduced as the holonomy of the connection.

We now come to the core of the theory of causal fermion systems: the {\em{causal action principle}}.
In order to single out the physically admissible
causal fermion systems, one must formulate constraints in the form of physical equations. To this end, we require that
the measure~$\rho$ be a minimizer of the causal action~${\mathcal{S}}$ defined by
\begin{align}
\text{\em{Lagrangian:}} && \L(x,y) &= \frac{1}{4n} \sum_{i,j=1}^{2n} \big(
|\lambda^{xy}_i| - |\lambda^{xy}_j| \big)^2 \label{Lagrange} \\
\text{\em{causal action:}} && \Sact(\rho) &= \iint_{\F \times \F} \L(x,y)\: d\rho(x)\, d\rho(y) \label{Sdef}
\end{align}
under suitable constraints (for the detailed form of the constraints see~\cite[Section~1.1.1]{cfs}).
A minimizer satisfies the {\em{Euler-Lagrange (EL) equations}}
which state that for a suitable parameters~$\s, \kappa>0$
(which can be understood as Lagrange multipliers of the constraints),
the function~$\ell : \F \rightarrow \R_0^+$ defined by
\beq \label{elldef}
\ell(x) := \int_M \L(x,y)\: d\rho(y) + \kappa \int_M \bigg( \sum_{j=1}^{2n} \big|\lambda^{xy}_j \big| \bigg)^2
\: d\rho(y) - \s
\eeq
is minimal and vanishes in spacetime, i.e.,\
\beq \label{EL}
\ell|_M \equiv \inf_\F \ell = 0 \:.
\eeq
These equations describe the dynamics of the causal fermion system.
In a suitable limiting case, referred to as the {\em{continuum limit}}, the EL equations
give rise to the equations of motion for the fields in the standard model and GR. The continuum limit consists in specifying the fermionic vacuum configuration as well as choosing a regularization which is admissible by the causal action principle. This setup is then studied in the limiting case when the regularization length is much smaller than all other length scales in the system. We will introduce the regularization in some detail below. The precise definition of the continuum limit as well as all the choices required in its construction has been worked out in~\cite{cfs}. 

We finally give more details on the notion of causality. 
In Definition~\ref{def:causalstructure} we already introduce notions of timelike, spacelike
and lightlike separation.
The Lagrangian~\eqref{Lagrange} is compatible with this notion of causality in the
following sense. Suppose that two points~$x, y \in \F$ are spacelike separated.
Then the eigenvalues~$\lambda^{xy}_i$ all have the same absolute value,
implying that the Lagrangian~\eqref{Lagrange} vanishes. 
Thus pairs of points with spacelike separation do not enter the action.
Even more, these pairs of points also drop out of the
EL equations~\eqref{EL}.
This can be seen in analogy to the usual notion of causality where
points with spacelike separation cannot influence each other.

In addition to the above notions of causality, a
causal fermion system also distinguishes a {\em{direction of time}}. 
Indeed, we can introduce the functional
\[ {\mathscr{C}} :  M \times M \rightarrow \R\:,\qquad
{\mathscr{C}}(x, y) := i \,\tr \big( y\,x \,\pi_y\, \pi_x - x\,y\,\pi_x \,\pi_y \big) \:, \]
which leads to the following notion of time direction (for details see~\cite[Section~1.1.2]{cfs}).
\begin{Def}\label{def:order}
For timelike separated points~$x,y \in M$,
\[ \left\{ \begin{array}{cll}
\text{$y$ lies in the {\bf{future}} of~$x$} &\!&  {\mbox{if~${\mathscr{C}}(x,y)>0$}} \\[0.4em]
\text{$y$ lies in the {\bf{past}} of~$x$} && {\mbox{if~${\mathscr{C}}(x,y)<0$}}
\end{array} \right. \]
\end{Def} \noindent
We finally point out that this notion of ``lies in the future of'' is not necessarily transitive.
This corresponds to our physical conception that the transitivity of the causal relations
could be violated both on the cosmological scale (there might be closed timelike curves)
and on the microscopic scale (there seems no compelling reason why the causal
relations should be transitive down to the Planck scale).
However, in~\cite{linhyp, localize} causal cone structures were constructed,
which do give rise to transitive causal relations (for details see in particular~\cite[Section~4.1]{linhyp}).
All these causal relations coincide on length scales which are much larger than the Planck length.
On very small length scales, however, the causal structure of spacetime may be more intricate,
making it necessary to work with different notions. A concrete exploration on how the causal structure
may look like on the Planck scale is given in~\cite{reg} (see in particular Figures~5-7).

In order to illustrate the concepts introduced in this section in a specific example, we will now discuss Minkowski space as a CFS. 

\subsubsection{Causal Fermion Systems in Minkowski Space}\label{sec:continuum}
In this section we explain using the example of the Minkowski vacuum how the abstract ideas introduced above are implemented in practice. 
For notational simplicity, we work in a fixed reference frame and identify Minkowski
space with~$\R^{1,3}$, endowed with the standard Minkowski inner product with signature
convention~$(+,-,-,-)$. We consider the smooth, spatially compact solutions of the
Dirac equation in Minkowski space
\[ 
    (i \gamma^k \partial_k - m) \,\psi = 0 \:, \]
endowed with the usual scalar product 
\[ ( \psi\, | \,\phi ) := \int_{t=\text{const}} (\overline{\psi} \gamma^0 \phi)(t, \vec{\x})\, d\vec{\x} \:, \]
where~$\overline{\psi} = \psi^\dagger\gamma^0$ is the adjoint spinor and $\vec{\x}$ the elements of the $t=\text{const}$ hypersurface. We then choose the Hilbert space~$\H$ to be the completion of the subspace of all negative-energy solutions.
We denote the restriction of the scalar product by~$\la .|. \ra := (.|.)|_{\H \times \H}$.

We now want to build the local correlation map assigning an operator~$F(\x)$ for any point~$\x$ of Minkowski space. This is obtained by applying Riesz`s representation theorem to the bi-linear map
\[ b_\x(\psi, \phi) = -\overline{\psi(\x)} \, \phi(\x)  \qquad \forall \psi, \phi \in \H \:. \]
We thus obtain the map to be
\begin{equation}\label{eq:correlation}
    \la \psi\, |\, {F(\x)}\, \phi \ra = -\overline{\psi(\x)} \, \phi(\x)  \qquad \forall \psi, \phi \in \H.
\end{equation}
This operator encodes information on the local densities and correlations of all the wave functions in~$\H$
at the classical spacetime point~$\x$. Therefore, it is referred to as the {\em{local correlation operator}}.
By construction, this operator is self-adjoint, has rank at most four and has
at most two positive and at most two negative eigenvalues.
Therefore, following Definition~\ref{def:cfs}, the local correlation operator~$F(\x)$ is an element of~$\F$
if we choose the spin dimension~$n=2$.
Hence we have successfully constructed the local correlation map~$F[\eta_{\mu\nu}]$
from Minkowski vacuum to~$\F$. According to the construction~\eqref{eq:pushforward} this gives us the measure 
\beq \label{pushforward}
\rho = F_* \mu \qquad \text{or equivalently} \qquad
    \rho(\Omega) := 
    \mu \big( {F^{-1}(\Omega)} \big) \:,
\eeq
where~$\mu$ is the standard Lebesgue measure on Minkowski. We thus have constructed a CFS~$(\F, \H, \rho)$ of spin dimension two.

Now, the above explanation was a bit oversimplified because the wave functions in~$\H$
are defined only up to sets of measure zero, and therefore the right side of~\eqref{eq:correlation}
is generally ill-defined point-wise. As a consequence we need to introduce a regularization by setting
\[ 
\psi_\varepsilon={\mathfrak{R}}_\varepsilon(\psi) \:, \]
where the {\em{regularization operator}}~${\mathfrak{R}}_\varepsilon : \H \rightarrow C^0(\R^{1,3}, \C^4) \cap \H$
is a linear operator which maps to continuous Dirac solutions\footnote{Here, for simplicity of presentation, we have chosen a regularization which is compatible with the Dirac equation. This need not be the case for general regularizations in general spacetimes.} and fulfills the relation \footnote{A simple example for a regularization is the convolution by a suitable mollifier (for details see~\cite[Example~1.2.4]{cfs}).}
\[ \psi=\lim_{\varepsilon \searrow 0}{\mathfrak{R}}_\varepsilon(\psi). \]
Working with the regularized wave functions, the right side of~\eqref{eq:correlation}
is a bounded sesquilinear form. Therefore, we can introduce the
{\em{regularized local correlation operator}}~$F^\varepsilon(\x)$ by
\[ 
    \la \psi_\varepsilon \,|\, {F^\varepsilon(\x)} \,\phi_\varepsilon \ra := -\overline{\psi_\varepsilon(\x)} \, \phi_\varepsilon(\x)  \qquad \forall \, \psi, \phi \in \H \]
and applying the above construction to the regularized local correlation map~$F^\varepsilon$ gives a CFS~$(\H, \F, \rho_\varepsilon)$.

It is shown in~\cite[Section~1.2]{cfs}, that all the structures of Minkowski space can be recovered
from this CFS in the limit~$\varepsilon \searrow 0$.
For example, in this limit the causal structure of Definitions~\ref{def:causalstructure} and~\ref{def:order},
reproduces the causal structure of Minkowski space and the connection~\eqref{eq:cfsconnection} agrees with the spin connection on Minkowski space. 

It turns out that to describe Minkowski vacuum we have to choose~$\H$ as the space of all negative-energy solutions of the Dirac equation. This in a sense realizes Dirac's original concept of the {\em{Dirac sea}}, that in vacuum, all the states of negative energy should be occupied. In the theory of causal fermion systems, the Dirac sea arises as the realization of the Hilbert space~$\H$ in terms of physical wave functions. Therefore CFS does not share the problems of the original concept (like the infinite negative energy density of the sea) because the Dirac sea, describing the underlying Hilbert space, drops out of the Euler-Lagrange equations derived from the causal action principle\footnote{The attentive reader might note, that this problem of the infinities originating in the Dirac sea (and many more infinities with different origin) has already been solved in QFT with its associated toolbox of regularization and renormalization flow. The point is, to make progress, CFS first takes one step back, reinstating the Dirac sea in its original form. CFS then solves the problem in a different way, which eventually allows for the re-derivation of QFT, but now with gravity built in from the start. }. 

The intuition for the action of the regularization operator is, that it ``smoothens'' the wave functions on a microscopic scale. In the theory of causal fermion systems, the regularization is not merely a technical tool in order to make divergent expressions finite, but it realizes the concept that on microscopic length scales, the structure of spacetime itself is modified. Thus we always consider the regularized objects as the physical objects.

In order to describe systems involving particles and/or anti-particles, following Dirac's hole theory one extends~$\H$ by Dirac solutions of positive energy and/or removes vectors of negative energy. Bosonic fields (like electromagnetic or gravitational fields), on the other hand, correspond to collective ``excitations'' of the Dirac sea, where the particle wave functions satisfy the Dirac equation modified by a potential~${{\mathcal{B}}}$,
\[ (i \Pdd + {{\mathcal{B}}} - m ) \, \psi = 0 \:. \]
If we build the local correlation map~$F \,[\eta_{\mu\nu}, B_\mu]$ from these solutions, varying the field~$B_\mu$ induces a variation of the measure constructed by~\eqref{eq:pushforward}. This measure can again be analyzed in the limit~$\varepsilon \searrow 0$ to obtain the equations of motion for the vector potential~$B_\mu$. To do so we can use the fact that the Minkowski vacuum with~$B_\mu=0$ is a critical point of the causal action. We can therefore study the linearization of the Euler Lagrange equations around this minimizer. As already mentioned above, this analysis of the {\em{continuum limit}} makes it possible to derive classical field equations like the Maxwell and Einstein equations from the causal action principle (for details see~\cite{cfs}).

\subsection{Modified Measure Theories} \label{sec:nrmt}

MMT were originally motivated by the cosmological constant problem. On the mathematical level this boils down to an inconsistency between the matter sector and the gravitational sector. As it is well known, in non-gravitational physics, like in particle mechanics, for example,  the origin from which we measure energy is not important. In mathematical terms that means that the equations of motion are invariant under addition of a constant to the matter Lagrangian~$\mathcal{L}_m$
\[ \mathcal{L}_m \longrightarrow  \mathcal{L}_m+C \:. \]
The same does not hold true for the standard second order variational formulation of GR based on the Einstein Hilbert action
\[ S= \int_M R(g) \, \sqrt{-g} \, d^4\x \:. \]
If we add a constant to the Lagrangian, 
\[ \mathcal{L}_G \longrightarrow  \mathcal{L}_G+C \:, \]
the equations of motion get an extra contribution of the form~$C g_{\mu\nu}$ from the variation with respect to the metric~$g$ of the additional~$C \int_M \sqrt{-g}d^4\x$ term in the action. In the semi-classical setup this makes the gravitational equations dependent on the vacuum expectation value of all the quantum fields. To make these calculations compatible with observation requires a lot of fine tuning of the parameters in the matter Lagrangian.

MMT are one way to resolve this conceptual inconsistency between the matter and the gravitational sector. Instead of integrating the Lagrangian against the measure~$\sqrt{-g}d^4\x$ we pick the measure to be an independent quantity. The total action is then given by 
\begin{equation}
    S= \int_M L \, \Phi(A) \, d^4\x \qquad \text{ with } \qquad L=\frac{-1}{\kappa} \, R(\Gamma, g) +L_m,
    \label{ActionwithPhi}
\end{equation}
where~$\kappa$ is the gravitational coupling constant. The scalar curvature is given in terms of the connection and the metric and the measure by 
\[ \Phi(A)= \frac{1}{6} \:\varepsilon^{\alpha\beta\mu\nu}\partial_\alpha A_{\beta\mu\nu} \:, \]
where~$A_{\beta\mu\nu}$ is the tensor gauge potential of a non-singular exact~$4$-form~$\omega=dA$. Thus the modified measure density~$\Phi(A)$ is given by the scalar density of the dual field-strength associated with that potential. As a result,~$\Phi(A) d^4\x$ is invariant under general coordinate transformations. We would like to emphasize at this point, that MMT is an umbrella term for a whole class of physical models with the common feature being, that the general measure, introduced above, replaces the metric measure for the integration in the variation principle.  

This approach to modify gravity was first introduced in~\cite{NGVE} and later expanded upon in~\cite{scaleinvtmt} to include applications concerning spontaneous symmetry breaking of scale invariance. Field theoretic aspects were discussed in~\cite{Fieltheory}. After developing some examples using only one measure, multi-measure theories were developed, for example see~\cite{manymeasures} where it was demonstrated that those can accommodate cosmological solutions with an early inflationary phase and a late time exponential growth phase with a small cosmological constant. Modified Measures have also been used in the formulation of string and brane theories with dynamical tensions~\cite{stringsandmeasures}. For a recent review on the applications of Modified Measures to Modified Gravity Theories and Cosmology see~\cite{measuresModifiedGravityandCosmology}. 

Finally, we notice the non-holomorphic structure of standard GR due to the use of the measure $  {\sqrt{-g}}$. This problem is avoided when using a modified measure while still introducing the determinant $ {-g}$ into the action. In this way we can construct 
 complex extensions of GR, or other gravity theories as holomorphic MMTs. For more details see~\cite{HolomorphicMTM}.

\subsubsection{Single Measure Theory}\label{sec:singlemeasure}
Among the papers using modified measure  using only one measure, we can emphasize~\cite{NGVE}, 
 which was in fact the first paper on modified measures. Later generalizations allowed to include gauge fields, like in~\cite{GravityCosmologyandParticlePhysicswithouttheCosmologicalConstantProblem}, \cite{Hehl} and~\cite{spacefillingbranesandhigherdimensions}.  Also string theories with a modified measure have been formulated, see~\cite{stringsandmeasures} and follow up papers.

In two dimensions, where Einstein gravity becomes trivial because the Einstein tensor vanishes identically, a single modified measure provides an acceptable theory of gravity. That is, we can consider  
\[ S= \int_M  R(\Gamma, g) \,\Phi\, d^2\x \:, \]
where
\[ \Phi(A)= \frac{1}{6}\,\varepsilon^{\alpha\beta}\partial_\alpha A_{\beta} \:, \]
and where the variation with respect to the gauge field~$ A_{\beta}$ leads to the condition that the
curvature~$R = M$, where~$M$ is a constant. This is in fact the Jackiw-Teitelboim model~\cite{JT}.

Returning to~$3+1$ dimensions let us recall the difference between the first order, or Palatini, formalism and the second order formalism. In the first order formalism the connection~$\Gamma$ is considered as an independent degree of freedom. In the second order formalism it is given by the Levi-Civita connection. The name formalism is deceptive here because generally
these are different theories with the single exception of Einstein gravity
\[ S= \int_M  R(\Gamma, g) \, \sqrt{-g}\,d^4\x \:, \]
where the two formalism give the same equations of motion. When the gravitational coupling in the Lagrangian differs from that in the Einstein-Hilbert action, e.g., in $f(R)$ gravity, when higher curvature terms are included or when one works with modified measures, then they are different theories. There is no purely theoretical way  to determine which is best, only which one fits better with experiment and hence we will discuss MMT in both formalisms. 

For the first order formalism, following~\cite{NGVE}, if we vary the action~\eqref{ActionwithPhi} with respect to  the measure field~$A_{\beta \mu\nu}$, we get that the derivative of the Lagrangian is zero. This implies that the following equation is satisfied, 
\[ \frac{-1}{\kappa} R(\Gamma, g) +L_m= M \:, \]
where~$M$ is the constant of integration. For the first order formalism that is the end of the story.  The measure does not introduce new degrees of freedom, instead it is solved in terms of other fields through a constraint equation. 
The constant of integration~$M$ corresponds to the cosmological constant and has to be determined by observation.

In the second order formalism however we get more. Here it is convenient to introduce the scalar field~$\chi=\frac{\Phi(A)}{\sqrt{-g}}$. 
Following~\cite{NGVE} we get the following system of equations
\begin{align*}
    R_{\mu\nu}-\frac{1}{2}R \,g_{\mu\nu}= \frac{\kappa}{2}\left(T_{\mu\nu}+ M g_{\mu\nu}\right) + \frac{1}{\chi}\left( \chi_{,\mu;\nu} -\right.&\left.g_{\mu\nu} \, \Box\,\chi  \right) \\
   \Box \, \chi- \frac{\kappa}{D-1}\left[ \left(M + \frac{1}{2} \, T\right)+ \frac{(D-2)}{2}\,L_m           \right]&\,\chi =0
\end{align*}
Here, $T$ is the trace of the stress energy tensor and the integration constant~$M=-\Lambda$ takes the role of the cosmological constant. We see that this system of equations can only give rise to solutions compatible with the equations of Einstein's GR if~$\left(M + \frac{1}{2} T\right)+ \frac{(D-2)}{2}L_m    =0$. This condition is in particular satisfied by a classical Minkowski vacuum spacetime. However, it fails in classical de Sitter spacetimes. Therefore classical de Sitter spacetimes are not a solution to modified measure theories with a single measure in the second order formulation.

An interesting  consequence of the above results in the second order formalism is that one gets a non-conservation of the conventionally defined matter stress energy, that does not include a contribution from the field~$\chi$ , 
\begin{equation}
     \label{eq:nonconservation}
    T_{\mu\nu}^{\phantom{\mu\nu};\mu}= -2\, \frac{\partial \mathcal{L}_m}{\partial g^{\mu\nu}}\,g^{\mu\alpha}\nabla_\alpha \ln{\chi} \:. 
\end{equation}
The non-conservation is dependent on the gradient of the scalar field~$\chi$. That means, in a sense we can transfer ``energy'' from the measure, hence the gravitational sector of the theory, to the matter sector of the theory. 

In~\cite{Guendelman:1999qt} the so-called Einstein frame was introduced to examine the physical implications of the theory When working with an action of the type defined by~\eqref{ActionwithPhi} , to get to the Einstein frame one applies a local conformal transformation
\begin{equation}
    \label{EinsteinFrame}
g_{\mu\nu}\longrightarrow \bar{g}_{\mu\nu} =\chi \, g_{\mu\nu} \:,
\end{equation} 
which restores the gravitational equations to the  Einstein form. All additional equations are expressed just as well in terms of the metric~$\bar{g}_{\mu\nu}$.

In the first order formalism, the scalar field ~$\chi$ is not dynamical,does not introduce new degrees of freedom and can be solved for. Accordingly, when changing to the Einstein frame it only rearranges the interactions when reintroduced into the equations of motion. To some extent this reformulation implies running coupling constants in the Einstein frame. This is due to the fact that when performing the conformal transformation~\eqref{EinsteinFrame}, terms in the Lagrangian that are not conformally invariant get a~$\chi$-dependent weight. For a general discussion see~\cite{FOUNDATION}. In the case of the standard model all interactions are dimensionless, except for the Higgs field's potential terms.
 
In the first order formalism, the Einstein frame has another advantage. In the Hamiltonian formulation of the theory the original metric does not have a canonically conjugated momentum. In contrast, the canonically conjugated momentum to the connection turns out to be a function exclusively of the Einstein frame metric. Following a very similar calculation, in the case of general relativity, e.g., in the book by Sundermeyer  \cite{Sundermeyer}  and references therein, using the action defined in  \ref{ActionwithPhi} the canonically conjugate momenta to $\Gamma^\lambda_{\mu  \nu}$ becomes,

\begin{equation}
    \label{momenta of connection}
\pi^{\mu  \nu}_\lambda = \frac{-1}{\kappa} \Phi (g^{\mu  \nu}\delta^0_\lambda -\frac{1}{2}g^{\mu 0}\delta^\nu_\lambda -\frac{1}{2}g^{\nu 0}\delta^\mu_\lambda )  = \frac{-1}{\kappa} \sqrt{-\bar{g}} (\bar{g}^{\mu  \nu}\delta^0_\lambda -\frac{1}{2}\bar{g}^{\mu 0}\delta^\nu_\lambda -\frac{1}{2}\bar{g}^{\nu 0}\delta^\mu_\lambda )
\end{equation} 

The Einstein metric is therefore a genuine dynamical
canonical variable, as opposed to the original metric, which has zero canonical momentum. Therefore only the Einstein metric appears in the Hamiltonian. In this sense, going to the Einstein frame is equivalent to transitioning from the Lagrangian formulation of the theory to the Hamiltonian formulation.

We now briefly return to the second order formulation of the theory, where the conformal factor $\chi$ in the transformation to the Einstein frame corresponds to a dynamical scalar field. In this case going to the Einstein frame and rewriting the field as~$\chi= \exp{u/\sqrt{3}}$ leads to a  standard kinetic term in the action for the field~$u$~\cite{measuresModifiedGravityandCosmology}. This scalar field represents a new degree of freedom and the constraint equation becomes a dynamical equation for this field. 
Because the Einstein tensor of the re-scaled metric is divergence free, a total conserved energy momentum can be defined. However, it contains a contribution from the scalar field~$u$ in addition to the matter.

\subsubsection{Multi-Measure Theories}
Instead of action principles with a single measure we can also consider action principles of the form
\begin{equation}
    \sum_{i=1}^n \int L_i\Phi(A_i) d^4\x,
\end{equation}
where the Lagrangian $L_i$ is integrated with respect to the measure density $\Phi(A_i)$. Such multi-measure theories can be formulated for both the first order or the second order formalism. Most work on multi-measure theories has been done for the case of TMT where we have $n=2$.

In a recent study~\cite{measuresandfullhistoryofuniverse} the application of TMT in the first order formalism to all phases of the universe, stating from a non singular emergent phase, followed by inflation and then by a Dark matter and Dark Energy era has been investigated. Here, the fact that the scalar field~$\chi$ rearranges the interactions when reintroduced into the equations of motion plays a key role. The rearrangement manifests, for example, in the masses and interactions of the matter fields as well as in the unusual structure of their contributions to the energy-momentum tensor.
All these quantities appear to be~$\chi$  dependent which leads to the appearance of unsuspected interactions.

The result in the Einstein frame is an effective Lagrangian containing effective scalar field potentials that are non trivial functions of the integration constants and the original potentials that coupled to the measures~$\Phi$
and~$\sqrt{-g}$ .
In the context of cosmology, for example in ~\cite{GUENDELMANKATZ}, this can lead to a scalar field potential with two flat regions. One flat region is used to describe inflation and the other is used to describe the slowly accelerated phase of the universe observed now.

Furthermore, it has been shown that the well known theory of unimodular gravity when formulated in a generally covariant form~\cite{UnimodularGRI} 
\[ \mathcal{S}= \int d^4\x \,\sqrt{-g} \, (R + 2 \Lambda + \mathcal{L}_m) - \int d^4 \x \, \Phi(A) \,2 \Lambda \:, \]
appears as a special case of TMT. Note that in this case the first and second order formalism are indeed the same theory, as the Ricci scalar couples linearly to the metric measure as in the Einstein-Hilbert action. 
Here, a priori, $\Lambda$ is a dynamical scalar field, and the measure density~$\Phi(A)$ is as above.
Variations with respect to the potential~$A$ imply that the scalar field~$\Lambda$ is a constant, whereas variations with respect to the scalar field~$\Lambda$ yield~$\Phi(A) =\sqrt{-g}$. 

Another interesting example are TMT models that can relate to the mimetic theory of gravity~\cite{chamseddine2013mimetic} in the cosmological setting. Instead of forcing the Lagrangian to equal a certain constant through the use of a Lagrange multiplier, we will show that similar results can be obtained via the use of modified measures.
For that we use the following non-conventional gravity-scalar-field
action~\cite{guendelmantwo}  as a starting point:
\begin{equation}
S = \int d^4 \x \sqrt{-g}\, R + \int d^4 \x \bigl(\sqrt{-g}+\Phi(A)\bigr) L(\phi,X) \; .
\label{TMTDEDM}
\end{equation}
Here
$L(\phi,X)$ is the general-coordinate invariant Lagrangian of a single scalar field 
$\phi (x)$ of a generic ``k-essence'' form 
\begin{equation*}
L(\phi,X) = \sum_{n=1}^N A_n (\phi) X^n - V(\phi) \; , \qquad
X =- g ^{\mu\nu}\partial_\mu \phi \partial_\nu \phi \; .
\end{equation*}
This was also generalized to a more generic form for $L$, for example see~\cite{guendelman2015dark}.
Varying w.r.t. $g^{\mu\nu}$, $\phi$ and $A_{\mu\nu\lambda}$ yield the
following equations of motion, respectively:
\begin{equation}
R_{\mu\nu} -\frac{1}{2} Rg_{\mu\nu} =  T_{\mu\nu}
\label{einstein-eqs}
\end{equation}
\begin{equation}
T_{\mu\nu} = g_{\mu\nu} L(\phi,X) + 
\Bigl( 1+\frac{\Phi(A)}{\sqrt{-g}}\Bigr) \frac{\partial{L}}{\partial{X}}\partial_\mu \phi\partial_\nu \phi \:.
\label{EM-tensor}
\end{equation}
The variation of the measure fields $A$ require  that $L = M = constant $, implying that the first term proportional to the metric behaves exactly like a cosmological constant, or Dark Energy. The second term has exactly the behavior of cosmic dust for a scalar field that depends on cosmic time. These results are again independent of the formalism.

\section{Comparison} \label{sec:comparison} 
In this section we will highlight and compare the mathematical structures and the conceptional ideas behind CFS and MMT.
\subsection{Foundations of the Approaches}\label{sec:foundations}
We begin with a discussion of the basic building blocks of the theories.

\subsubsection{MMT}
MMT are formulated on classical manifolds with the Lagrangian for both matter and gravitation unchanged. It assumes the measure to be independent of the metric. In the second order formalism it thereby introduces  one  additional scalar degree of freedom  not present in the standard formulation of GR and the standard model. Meanwhile in the first order formalism the measure, through the equations of motion,  can be resolved in terms of other fields. In this case the measure does not introduce new degrees of freedom, but just adds a constant of integration.

\subsubsection{CFS}
CFS is formulated on an operator manifold~$\F^{\text{reg}}$ over a Hilbert space and the measure is the only ``dynamical'' degree of freedom in the theory. However, the fact that the support of the measure is typically restricted to a proper, lower dimensional subset of the operator manifold together with the intrinsic causal structures allows for the measure to encode a plethora of physical systems. In contrast to GR, for example, the operator manifold on which the variational principle is formulated does not fix the topology of the spacetime described by a minimizing measure.

\subsubsection{Discussion}
Given the different starting points for the theories,  a comparison is only possible close to the continuum limit where the regularized local correlation map~$F_\varepsilon \,[g_{\mu\nu}, A_\mu, \dots]$ allows us to think of a CFS in terms of classical spacetime structures. It is immediately clear from the construction of the measure~\eqref{eq:pushforward} that, in this setup, the measure on the manifold does not have to coincide with the measure derived from the metric~$g_{\mu\nu}$ which entered the local correlation map~$F_\varepsilon[g_{\mu\nu}, A_\mu, \dots]$. 

\subsection{Splitting the Tasks of the Metric}\label{sec:metric}
In the standard second order formulation of GR the metric encodes all gravitational degrees of freedom in the Einstein-Hilbert action. In particular the metric encodes 
\begin{enumerate}
    \item the map from the tangent space to the co-tangent space,
    \item the causal structure,
    \item the connection, 
    \item and the measure. 
\end{enumerate}
Furthermore, the action is composed from the matter fields and the metric, whereby the gravitational part of the action is exclusively composed from the metric.
It is plausible that in a fundamental theory not all of these properties will be encoded by the same mathematical object and that additional degrees of freedom are introduced by such a cut. 

First a comment on the standard formulations of gravity for comparison. In the first order formulation the connection is promoted to an independent field that appears in the action in combination with the metric. It turns out that for pure gravity this is equivalent to the second order formulation. As already mentioned in Section~\ref{sec:nrmt}, this equivalence is not true in a more general setting. For example, when matter is added to the model or, when the action contains higher order curvature terms. In these cases, by making the connection independent of the metric, we obtain new gravitational theories.

\subsubsection{MMT}
It is well known, that geometry is given by the combination of causal structure plus volume, i.e., measure density (see~\cite{hawking1976new,malament1977class} and, e.g.,\cite[p.9]{surya2019causal} for a discussion of this fact in the context of Causal Sets Theory). With that in mind it is quite a natural cut to consider the measure as an independent degree of freedom as is the case in MMT. In the second order formalism the metric still encodes (1)-(3) and the measure becomes an independent dynamic degree of freedom encoded in the field~$\chi$. The gravitational action is now composed of the metric and the measure density.In the first order formulation however the metric now only encodes (1) and (2) while the connection and the measure are independent variables. In this case the field~$\chi$ can be resolved in terms of other fields. The gravitational action is now composed of the metric, the connection and the measure. As discussed in Section~\ref{sec:singlemeasure} in contrast to GR this now leads to different equations of motion from the second order formulation. Being different theories, ultimately only experiment/observation can decide which formalism should be preferred. Conceptually, one might prefer the first order formalism based on the fact it already makes one fundamental geometrical object independent of the metric and MMT then goes one step further. In this sense, if one buys into the modified measure program, it might seem more natural to use the first order formalism.

\subsubsection{CFS}
In CFS the causal structure and the connection is already encoded in the operator manifold~$\F^{\text{reg}}$: the causal structure via the eigenvalues of the operator product~$xy$ between any two points in the
manifold (see Definition~\ref{def:causalstructure}) and the connection \eqref{eq:cfsconnection} via the polar decomposition of the fermionic projector~$\pi_y x$ mapping from~$S_x$ to~$S_y$. Given the fact that~$S_x$ is a subspace of a Hilbert space, it is itself a Hilbert space with the induced scalar product.  
In the limiting case the space~$S_x$ in some sense corresponds to the fiber at~$\x$ and we obtain the bundle as the emergent limiting structure. 

In CFS the measure~$\rho$ plays the role of the fundamental degree of freedom encoding the geometry and the matter content of the limiting classical spacetime. Again, it is crucial to note that the support of the measure varies for different minimizers, thereby encoding physical systems with a different causal structure and ultimately even a different topology of the limiting classical spacetimes. 

The map between the tangent and the co-tangent space is obtained in CFS as follows. First of all, for the tangent and co-tangent space to be well-defined, we need to assume that the spacetime~$M:= \supp \rho$ is a four-dimensional smooth manifold. Under this and a few technical assumptions (more precisely, that the tangent cone measure be non-degenerate; see~\cite[Definition~6.7]{topology}), it is shown in~\cite[Section~6]{topology} with measure-theoretic methods that at every spacetime point~$x \in M$ there is a distinguished Clifford subspace~${\mathscr{C}}\ell_x$ together with a canonical mapping~$\gamma_x$ from the tangent space~$T_x M$ to the Clifford subspace,
\[ \gamma_x \::\: T_x M \rightarrow {\mathscr{C}}\ell_x \:. \]
This mapping corresponds to the usual Clifford multiplication. The anti-commutation relations~$\{ \gamma_x(u), \gamma_x(v)\} =: 2 g(u,v)$ define a Lorentzian metric, which gives rise to the usual identification of the tangent and co-tangent space.

The Lagrangian is a bi-distribution~$\mathcal{L}(x,y)$ built from the eigenvalues of the operator product of the two spacetime points~$x,y \in \F$ and integrated against the measure over both variables~$d\rho(x)\,d\rho(y)$ to obtain the causal action. Only the measure is varied to find a minimizer of the action.

\subsubsection{Discussion} 
The first order formulation promotes one of the functions of the metric in the second order formulation to an independent variable, namely the connection. MMT then makes an additional cut to either the first or the second order formalism by separating the measure from the metric. 
CFS on the other hand is much more radical, promoting the measure to be the only fundamental degree of freedom while the connection and the causal structure are already encoded in the operator manifold~$\F^{\text{reg}}$. 

At face value, the cut made by the first order formalism does not fit well conceptually with CFS as there both the causal structure and the connection are encoded in similar mathematical objects, namely the eigenvalues of the operator product and the polar decomposition of the fermionic projector. However, those are not really variables in the CFS theory therefore there might as well be a cut in the effective description. 

Finally, CFS only features a single fundamental measure making it seem unlikely that a multi-measure theory could emerge as an effective description.

\subsection{Derivation of the Field Equations}\label{sec:eom}
Both MMT as well as CFS in the continuum limit, are field theories based on a variation principle. Ultimately, the predictive power of any such theory lies in its field equations. Here we discuss how the field equations are obtained. 

\subsubsection{MMT}
The variation principle in MMT is identical to standard GR except for the fact that we have an additional variable, the measure. Accordingly to obtain the full set of field equations we also have to vary the action with respect to this variable. While for theories, where the gravitational sector is described by the Einstein-Hilbert action, the predictions of the first order formulation and second order formulation coincide, this is not true anymore for MMT. The field~$\chi$ only becomes a dynamical scalar degree of freedom in the second order formulation. 

\subsubsection{CFS}

A priori, in CFS there is only a single object, the measure~$\rho$, to be varied in the action principle. However, close to the continuum limit, where we can approximate the CFS by standard fields in a classical spacetime, we can vary the measure in a more controlled fashion according to the effective degrees of freedom in the continuum description. This is achieved by varying the fields that enter the local correlation map~$F_\varepsilon \,[g_{\mu\nu}, A_\mu, \dots]$. The standard results for matter and gravitational field equations are obtained in the continuum limit
where the regularization length is taken to zero~$\varepsilon\searrow0$. However, given the fact that matter only couples to gravity as a third order effect of an expansion in the regularization length~$\varepsilon$, we know that for CFS to describe the physical world we live in, the regularization length has to be finite~$\varepsilon\neq 0$. Accordingly, one expects deviations from the standard results in the effective description. This was demonstrated as a proof of principle in the case of the CFS mechanism for baryogenesis~\cite{baryogenesis} and requires modified measures for a minimizing classical spacetime and matter configuration as we will elaborate now.

The method described in Section~\ref{sec:continuum} for constructing a causal fermion system in Minkowski space
generalizes in a straightforward way to curved classical spacetimes (for details see~\cite[Section~1]{nrstg}).
However, analyzing the causal action principle in curved classical spacetimes yields surprising effects.
We now outline those findings which are of relevance for the comparison with MMT.

To set the stage, let~$(\scrM, g)$ be a classical globally hyperbolic spacetime.
Choosing a subspace~$\H$ of the Dirac solution space as well as a regularization
operator~${\mathfrak{R}}_\varepsilon : \H \rightarrow C^0(\scrM, S\scrM) \cap \H$
(where~$S\scrM$ denotes the spinor bundle), we define the local correlation operator
similar to~\eqref{eq:correlation} by~$\la \psi\, |\, {F(\x)}\, \phi \ra = -\Sl \psi(\x) | \phi(\x) \Sr_\x$
(where~$\Sl .|. \Sr_\x$ is the inner product on the spinor space~$S_\x\scrM$).
Following the procedure in~\eqref{pushforward}, we can take the push-forward measure of the
volume measure~$d\mu_\scrM := \sqrt{-\det g} \, d^4x$ in classical spacetime, i.e.,\
\beq \label{pushvolume}
\rho_\varepsilon := F_* (\mu_\scrM) \:.
\eeq
This gives a causal fermion system describing the curved classical spacetime~$(\scrM, g)$. 

The first observation is that the regularization behaves dynamically.
This means that, even if we choose a simple regularization on an initial Cauchy surface
(for example, a cutoff in momentum space with respect to an observer),
the dynamics as described by the Dirac equation
in curved classical spacetimes has the effect that, at a later time, the regularization has a more complicated
structure. In particular, the regularization length~$\varepsilon$ must not to be
considered as a constant (say, the present day Planck length), but instead it becomes a function~$\varepsilon(x)$
which changes dynamically on cosmological scales. This effect was worked out in detail in~\cite{reghadamard}
by introducing the so-called regularized Hadamard expansion. In this formulation, the dynamics of the
regularization is obtained by solving transport equations along null geodesics.

The second observation is that, in curved classical spacetime, the measure~$\rho$
constructed by taking the push-forward of the volume measure~\eqref{pushvolume} doesn't have to be a minimizer of the causal action principle anymore. Instead, one has to start with a more general measure of the form
\beq \label{pushh}
\rho_\varepsilon := F_*\big( \chi \,\mu_\scrM)
\eeq
for a smooth function~$\chi \in C^\infty(\scrM, \R^+)$ which is not constant. I.e., we consider the push-forward of a modified measure $\Phi(A)=\chi \,\mu_\scrM$ in curved classical spacetime. 

In simple terms, this result can be understood as follows (for details see~\cite[Section~4]{baryogenesis}).
In order for our causal fermion system to satisfy the physical equations, we must
satisfy the EL equation~\eqref{EL}. Hence the function~$\ell$ defined by~\eqref{elldef} must
vanish identically in spacetime. As just explained, in curved classical spacetime
the regularization length~$\varepsilon(\x)$ is a function of the spacetime point. This function
enters the integrals in~\eqref{elldef} in a non-trivial way, meaning that, working with the geometric
measure~\eqref{pushvolume}, the first two summands in~\eqref{elldef}
depend on the function~$\varepsilon(\x)$, but the last summand does not. As a consequence, the EL equations will
be violated. In order to satisfy these equations, we must work with~\eqref{pushh}
and choose~$\chi$ in such as way as to compensate for the spacetime dependence of~$\varepsilon(\x)$.
In this way, the causal action principle forces us to modify the measure according to~\eqref{pushh}.

\subsubsection{Discussion} 
MMT is a relatively minor modification of general relativity, and its field equations are accordingly compatible with standard predictions in the local universe. However, there is no fundamental argument that allows to distinguish which formulation, first order, second order or multiple measures, is fundamental. Accordingly, when building phenomenological models, there are many free parameters to fit observation. CFS, on the other hand, comes with a single fixed variation principle. However, in this case working out the full phenomenology is a difficult task. As argued above CFS requires modified measures in the effective description when working in curved classical spacetimes. This was first noted in the proof of principle derivation of a new mechanism for baryogenesis~\cite{baryogenesis} from CFS. 
 Modified measures turn out to be crucial for the consistency of the approximate effective description. The departure from geometric measures implies a non-conservation of the matter stress-energy tensor in the second order formulation \eqref{eq:nonconservation}; a property necessary when describing baryogenesis as a transition from a vacuum configuration. In a sense, the mechanism transfers ``energy'' from the gravitational sector, to the matter sector.

\section{Outlook and Conclusion}\label{sec:conclusion}

In the present paper we have discussed the relationship between MMT and CFS. While it is clear that a connection exists, a substantial amount of work will be needed to constrain those MMT that are compatible with CFS.
Working out the precise dynamics of the scalar field~$\chi$ for a CFS in the continuum limit will be subject of future work. In particular it will be important to work out whether CFS gives rise to the dynamics of the first or the second order formulation of MMT or whether one obtains something more general. If deriving such constraints in a rigorous manner is successful, this will allow to study these MMT models as effective phenomenological models for CFS in cosmological/astrophysical settings without having to take the full CFS formalism into consideration. Meanwhile multi-measure theories at first glance seem incompatible with CFS. Ideally, this will be different for other approaches to unification/quantum gravity, which would provide an efficient pathway to derive phenomenological predictions that can distinguish these approaches from CFS.

On the conceptual level there are open questions concerning the scalar field~$\chi$ that appears in MMT. It is a priori not  clear how to interpret this field exactly. It is tightly connected to the volume of the classical spacetime and can play the role of the inflation field when using the second order formalism~\cite{NONRIEMANNIANVOLINFLATION}. However, it only encodes deviations from the geometric volume. Hence, if one were to quantize the modified measure~\cite{dzhunushaliev2022quantization} and with it the scalar field~$\chi$, then quanta of this field can not directly be interpreted as a quantum of volume. It is interesting to note that in~\cite{dzhunushaliev2022quantization} modified measures are related to a change in the fundamental length scale. This is very similar to the relation discovered in~\cite{baryogenesis} where the modification of the measure is related to a change in the regularization length.  These questions are subject of ongoing research and we hope to shed some light on them in our upcoming papers~\cite{cfstdqg, gravvariance}. 

The present comparison shows that in absence of a full theory of quantum gravity it is worthwhile to study the effect of separating different aspects of the functions the metric fulfills in the second order formulation of GR into independent variables/degrees of freedom more systematically. This gives a set of conceptually well motivated modifications of gravity which stand a chance to be emergent as an appropriate limit of an underlying theory of quantum gravity. In particular with respect to  possible phenomenology on cosmological scales. Having, e.g., a literature with a wide variety of MMT models available will help significantly to speed up the investigation of phenomenological implications of CFS once the connection can be made rigorous. Similar research efforts regarding other, well motivated, modifications of GR might have similar synergies with various approaches to Quantum Gravity.

\Thanks{{{\em{Acknowledgments:}}
We would like to thank the COST Action 18108 QGMM for facilitating the exchange between different approaches to Quantum Gravity. The idea for this paper was born after two of the authors met during the writing of the survey article~\cite{addazi2022quantum}.
We are also grateful to the referees for their thoughtful comments and their valuable feedback.

\providecommand{\bysame}{\leavevmode\hbox to3em{\hrulefill}\thinspace}
\providecommand{\MR}{\relax\ifhmode\unskip\space\fi MR }
\providecommand{\MRhref}[2]{%
  \href{http://www.ams.org/mathscinet-getitem?mr=#1}{#2}
}
\providecommand{\href}[2]{#2}

\end{document}